\documentclass[english,utf8,hyperref]{lni}
\usepackage[utf8]{inputenc}
\usepackage{caption}
\usepackage{booktabs}
\usepackage{url}

\graphicspath{{graphics/}}
\newcommand{\halmd}{\textit{HAL's MD package}}

\newcommand{\Cpp}{C\nolinebreak[4]\hspace{-.05em}\raisebox{.3ex}{\scriptsize ++}}
\newcommand{\code}[1]{\texttt{#1}}

\begin{document}

\title{Gaining Cross-Platform Parallelism for HAL's Molecular Dynamics Package using SYCL}
\author[
	V. Skoblin
	\and
	F. Höf{}ling
	\and
	S. Christgau
]{
	Viktor Skoblin\footnote{Zuse Institute Berlin, Supercomputing Department, Takustraße 7, 14195 Berlin, Germany, \email{{skoblin,christgau}@zib.de}}
	\and
	Felix Höf{}ling\textsuperscript{1,}\footnote{Freie Universität Berlin, Department of Mathematics and Computer Science, Arnimallee 6, 14195 Berlin, Germany}
	\and
	Steffen Christgau\textsuperscript{1}
}

\booktitle{PARS} 

\maketitle

\begin{abstract}
Molecular dynamics simulations are one of the methods in scientific computing that benefit from GPU acceleration.
For those devices, SYCL is a promising API for writing portable codes.
In this paper, we present the case study of \halmd{} that has been successfully migrated from CUDA to SYCL.
We describe the different strategies that we followed in the process of porting the code.
Following these strategies, we achieved code portability across major GPU vendors.
Depending on the actual kernels, both significant performance improvements and regressions are observed.
As a side effect of the migration process, we obtained impressing speedups also for execution on CPUs.
\end{abstract}


\section{Introduction}
Molecular dynamics (MD) simulations have become an indispensable methodology in classical statistical physics and related application fields such as materials research, nanofluidics, and soft and active matter.
%
Although a wide range of applications for MD exist, they can often be addressed by a single software framework, combining many-particle states and simple time stepping rules with additive forces obtained from problem-specific interaction laws.
Thus, a software framework for particle-based MD simulations can serve as a tool to study phenomena at a wide range of scales.
Technically, such a framework performs time-stepping integration of Hamiltonian equations of motion for a large set of interacting particles in a spatial domain with periodic boundary conditions \cite{AllenTildesley:Simulation};
typical system sizes in applications range from few thousand to few million particles.
The inherent parallelism of virtually every part of the MD algorithm makes it highly suitable for accelerator hardware; in particular, time-stepping integration and force computations can be performed independently for each particle.

One such framework is \halmd{}~\cite{Colberg2009, HALMD}, which was developed with a focus on parallel performance using GPUs.
The code employs the CUDA programming model and, thus, it is currently limited to Nvidia GPUs.
Opposite to that, the diversity in the GPU market has been increasing with AMD and Intel offering according hardware.
SYCL~\cite{std:sycl:2020r5} appears to be a promising API for having a single code base that can be used across the different vendors.
Therefore, porting of applications like \halmd{} to SYCL increases their sustainability.

In this paper, we present the strategies and challenges (Section~\ref{sec:migration_strategies}) that were employed during the migration of \halmd{} (Section~\ref{sec:halmd}) from CUDA to SYCL.
Those might be helpful for other code porting efforts.
Based on the successful code migration we also perform performance studies on GPUs from different vendors (Section~\ref{sec:perf_eval}) and observe both positive and negative outcomes with respect to performance.

%


\section{Related Work}
\label{sec:related_work}
A number of publications report case studies of porting applications to SYCL.
In many cases, the Intel DP\Cpp{} Compatibility Tool\footnote{\url{https://www.intel.com/content/www/us/en/developer/tools/oneapi/dpc-compatibility-tool.html}} was helpful in the migration process: only small changes had to be done to make the code compile and execute on the target platforms~\cite{DBLP:conf/ipps/ChristgauS20, DBLP:conf/iwbbio/CostanzoRSNP22, DBLP:conf/waccpd-ws/FortenberryT22, DBLP:conf/supercomputer/SakiotisAPRTZ23}.
In those examples, the achieved performance of the migrated version is close to the original one on the same hardware and a good performance portability~\cite{osti_1332474} was observed.
However, for some cases the performance was decreased, e.g.\ due to the specific implementation of SYCL reductions and the varying availability of functions in the compiler backends for the targeted Nvidia and Intel platforms~\cite{DBLP:conf/supercomputer/SakiotisAPRTZ23}.

A number of major MD simulation codes have been ported from traditional MPI-based parallelization to support some kind of accelerator hardware using different programming models.
In biomolecular modeling, codes such as AMBER \cite{Salomon-Ferrer:WCMS2013}, CHARMM \cite{Brooks:JCC2009}, GROMACS \cite{Abraham:S2015}, and NAMD \cite{Phillips:JCP2020} rely on one or several vendor-specific APIs, including CUDA in all cases and sometimes HIP. 
The successful SYCL port of the GROMACS package exhibits an overall performance drop of 10\% but, at the same time, the simple kernels run faster in the migrated version~\cite{Alekseenko_2021}.
LAMMPS~\cite{Thompson:CPC2022}, the leading package in materials research, offers partial support for multiple accelerator APIs at different levels of maturity, which incures additional maintainance efforts.
Similarly, ESPResSo~\cite{weik19a} has only a partial port to CUDA, which is complemented by the CUDA-parallel tool MDsuite~\cite{Tovey:JC2023} for post-processing of the data.
On the other hand, MD packages such as HOOMD-blue~\cite{Anderson:CMS2020}, OpenMM~\cite{Eastman:PCB2017}, RUMD~\cite{Bailey:SP2017}, and \halmd~\cite{Colberg2009, HALMD}, were developed from the ground up targeting CUDA-supporting GPUs.

\section{HAL's MD package}
\label{sec:halmd}
\halmd{} is a high-precision molecular dynamics package that was designed to support acceleration by Nvidia GPUs using CUDA \cite{Colberg2009}.
So far, it has been used mainly in statistical physics for research on inhomogeneous fluids; typical applications include studies on
liquid--vapor interfaces \cite{Hoefling:2015}, phase separation dynamics \cite{Roy:2016},
kinetics of methane capture \cite{Hoeft:Thesis},
cavitation in glasses \cite{Chaudhuri:2016},
friction and viscosity \cite{Straube:CP2020} and heat conduction \cite{EbrahimiViand:JCP2020}.

\paragraph{Code structure.}

The code itself is written in \Cpp{14} and essentially forms a collection of modules with a Lua scripting interface to enable the flexible set up of simulation scenarios by \emph{inter alia} connecting slot functions to signals (see below).
The \Cpp{} classes behind virtually every module carry template parameters for the spatial dimension (2 or 3) and the floating-point type (single, mixed precision, double) of the particle data arrays \cite{Colberg2009};
for the present purpose, we stick to three dimensions and single precision for GPU and double precision for CPU benchmarks.
The modules are organized in nested namespaces and are divided into top-level categories for the computation of interaction forces, the time-stepping integration of the equations of motion, the evaluation of observables, and others.

\begin{figure}
	\centering
	\includegraphics[width=\textwidth]{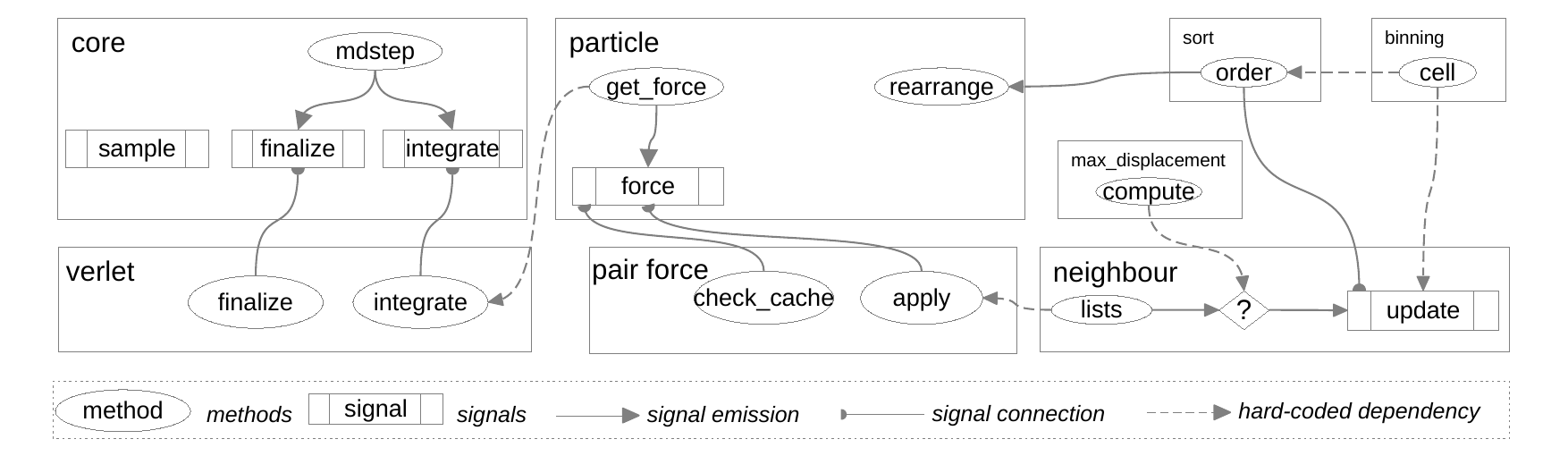}
	\caption{Signal--slots connections and data dependencies in a minimal MD simulation. Each box represents a module in \halmd, small rectangles represent signals, and ellipses refer to class methods, which may serve as slot functions.
	}
	\label{fig:signals}
\end{figure}

The software design follows a data-driven program flow:
The sampling of observables is the main objective and the corresponding evaluation methods are connected to the \code{sample} signal, which is emitted alternatingly with the signals for time integration (Fig.~\ref{fig:signals}).
The latter trigger the methods \code{integrate} and \code{finalize} of, e.g., the two half-steps of the velocity-Verlet integrator, which update particle momenta and positions.
A momentum update that follows a change of particle positions requires a new calculation of the interaction forces, which emits the \code{force} signal.
In general, the methods attached here make the most time-consuming part of the simulation.

In the case of pair interactions between all $N$ particles, the algorithmic complexity can be decreased from $O(N^2)$ to $O(N \log N)$ by truncating the range of the interaction potentials, which permits restricting the loop over all pairs to pairs of close particles.
To this end, one maintains a data structure called Verlet neighbor lists \cite{AllenTildesley:Simulation}, which are constructed efficiently using a spatial decomposition of the set of particles (``particle binning'').

\paragraph{Targeted Architectures.}
For each module in \halmd, two back-ends are available: an STL-focused version in pure \Cpp{} and one based on CUDA.
The former is a serial code for a single CPU core, which serves as a reference implementation;
however, it has always been maintained as an alternative code basis for future parallelization efforts.
The CUDA back-end makes use of the inherent parallelism of the MD algorithm by sequentially executing compute kernels that operate in parallel on the particle data arrays.
A CUDA implementation is available for all steps of the algorithms, which eliminates the need for data transfers between GPU device memory and host memory.
The implementation also carefully uses the different memory types of the Nvidia GPUs, like global, shared and texture memory.
The latter is also used to pack multiple data structures together which allows to load four floating point values at once. 

As a peculiarity of the CUDA implementations in \halmd, the kernel code are separated from the host code in different translation units.
Further, an abstraction layer (\code{cuda\_wrapper}) is used to hide the CUDA-specifics from the module implementation \cite{cuda-wrapper}.
The abstraction layer is built on-top of the low-level CUDA driver API and does not use the CUDA language extensions to C/\Cpp.
Kernels are launched by passing a function descriptor and the launch configuration as regular function arguments to \code{cuLaunchKernel}, thus avoiding the \code{kernel<<<launch\_config>>>} syntax.
It also permits the separate compilation of the host code with a regular \Cpp{} compiler, thereby avoiding the (decreasingly) limited language support of \code{nvcc}.
Moreover, \code{cuda\_wrapper} handles the allocation and copying of GPU device memory and fosters the RAII idiom.

\section{Code migration from \Cpp/CUDA to SYCL}
\label{sec:migration_strategies}

Given the two back-ends of \halmd, only the CUDA version provides parallelism, but is bound to a single GPU vendor.
Contrary, the pure \Cpp{} and therefore highly portable CPU reference implementation only runs sequentially, which is inefficient on contemporary processors.
Merging the upsides of both code bases is desirable and would result in a single code base that supports different platforms and lowers the maintenance requirements.
For this purpose, SYCL appears as an appropriate API, but different approaches have to be considered for the code migration as will be demonstrated in the following.

The code base includes about 4.3k lines of CUDA code in 73 files and a complete manual migration of the code would be tedious.
For the present study, we therefore focus on an all-to-all particle interaction initially, using the untruncated pair forces given by the common Lennard-Jones potential.
This is combined with the velocity-Verlet integrator and a stochastic, Anderson-like thermostat (termed \code{Boltzmann} thermostat in \halmd).
This choice was motivated by the practicality of porting the corresponding kernels from CUDA to SYCL; in particular,
the migration of more complex kernels related to the neighbor list construction can be bypassed at this stage.
In addition, the existing CUDA kernels all use aligned memory access and can therefore easily be migrated.
Their migration also allows a fair performance comparison against the native CUDA implementation.

%

\paragraph{Migration of CUDA code with Intel DP\Cpp{} Compatibility Tool.}
One option for the migration from CUDA to SYCL is to use the Intel's Compatibility Tool (or the \emph{SYCLomatic} tool); the process has already been described in the literature \cite{DBLP:conf/ipps/ChristgauS20, DBLP:conf/iwbbio/CostanzoRSNP22, DBLP:conf/waccpd-ws/FortenberryT22, DBLP:conf/supercomputer/SakiotisAPRTZ23}.
In case of \halmd, the automatic migration successfully migrated the targeted compute kernels (functions with the \code{\_\_global\_\_} keyword).
However, the tool failed to migrate the employed CUDA abstraction layer \code{cuda\_wrapper} \cite{cuda-wrapper} since it does not support the CUDA driver API.
Thus, all kernel launches were left untouched, which would entail considerable manual efforts to fix this.

Additionally, compute kernels and calling host code in \halmd{} are found in separate translation units, which conflicts with SYCL's single source model.
In SYCL, all captured data inside the kernel (lambda function), must be \emph{device copyable}~\cite[Section~3.13.1]{std:sycl:2020r5}.
The SYCL compiler employs heuristics to check if all captured objects inside a kernel are device copyable.
However, we noticed that both Intel's oneAPI dpcpp/icpx and the open source Intel LLVM-based compiler do not always detect correctly whether all objects actually satisfy this requirement.
For example, \Cpp{} functors are widely used in \halmd{}, e.g., to store potential parameters and call the potential evaluation from within the force kernels.
For a templated kernel function that is defined in a separate translation unit, declared with the \code{SYCL\_EXTERNAL} specifier, and which receives a functor as template parameter, this renders the whole kernel not being device copyable.

\paragraph{Migrating CPU Sequential Code to SYCL.}
Alternatively to the Compatibility Tool-assisted migration, it may be easier to port simple algorithms
to SYCL manually by adapting the original sequential code.
To this end, an existing CUDA version of a compute kernel may be used as a template for the SYCL implementation.
A similar approach was followed when porting GROMACS \cite{Alekseenko_2021}.

For this manual migration, possible data flows to and from the device need to be handled.
Our solution to a transparent handling of possible data transfers in SYCL for the migration efforts
was to add a double-buffering mechanism to the \code{particle} class of \halmd, which encapsulates all arrays storing particle properties.
Specifically, buffer arrays are pairwise allocated in device and host memory and are wrapped by a \code{halmd::cache} object, which allows tracking of read and write accesses.
Memory transfers are then performed to either buffer, and a synchronization is only done as needed.

Having established such a memory management scheme, the code can be migrated kernel by kernel while preserving the other parts' functionality.
For simple kernels, where each thread is assigned one particle and acts independently on the corresponding data, the migration amounts to simply replacing the \code{for} loop in the \Cpp{} code by SYCL's \code{parallel\_for} statement, generating a kernel invocation.
For optimal performance, only minor adjustments like setting the work group size were necessary.
With this approach, we have successfully migrated the set of modules needed for a basic MD simulation with an untruncated pair interaction (see Fig.~\ref{fig:signals}).
The employed strategy permitted a fast and straightforward migration, leading to clean and compact code without the structure being dictated by CUDA restrictions.

\paragraph{Mixed Strategy.}
In some cases, an algorithm can not be parallelized by only replacing loops as in the previous strategy.
In this case, we combined automatically migrated CUDA kernels and manually adjusted code (see previous paragraph).
This strategy also enabled the replacement of custom CUDA implementations of reduction and sorting by library calls.
For reduction, we employed SYCL's \emph{reduction variables}~\cite[4.9.2]{std:sycl:2020r5} and, for sorting, the template-based \emph{oneDPL} library\footnote{\url{https://spec.oneapi.io/versions/latest/elements/oneDPL/source/index.html}} from the oneAPI packages.
This approach also allowed for SYCL-specific code tunings, such as the usage of \code{sycl::event} to define kernel dependencies.

\paragraph{Migration of HAL's MD package}
\label{sec:halmd:migration}

Following the mixed and manual strategies enabled us to successfully migrate significant portions of \halmd{} from CUDA to SYCL.
This also included to migrate the non-trivial determination of the kernel launch configurations and kernel dependencies.
Nevertheless, the open source SYCL implementation by Intel has only limited and experimental support for textures (\code{sycl::image}).
For the migration, according code paths had to be changed to use global memory instead.

\section{Performance Evaluation}
\label{sec:perf_eval}
To assess the benefit of the SYCL migration efforts, we conducted performance experiments both on the CPU as well as on different GPUs.
We start with a setup that involves kernels for an all-to-all interaction, which were migrated manually and straight-forward in order to get a first impression of the achievable performance and portability.
Afterwards, we focus on the practically more relevant computation using truncated pair forces.

\begin{figure}
	\centering
	\includegraphics[width=\linewidth]{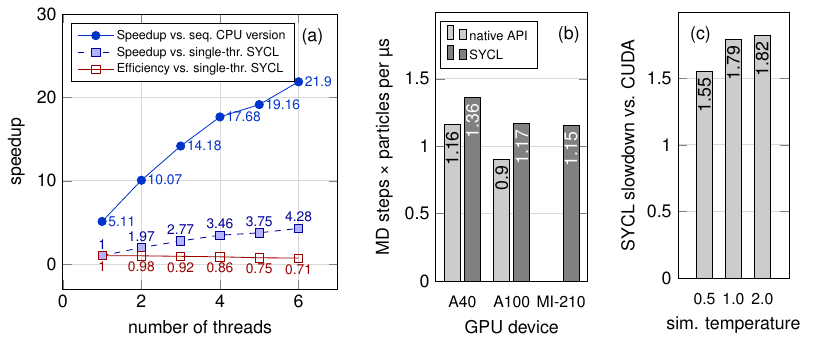}
\caption{Performance data for all-to-all interaction simulation.
~(a)~Speedups for the SYCL version on Intel Cascade Lake CPU compared to the sequential baseline and the SYCL version using a single thread, and parallel efficiency for the latter.
~(b)~Performance of the SYCL and native version on different GPUs (higher number is better) for all-to-all interaction.
~(c)~Slowdown of the SYCL version vs. CUDA for different simulation temperatures with truncated pair interaction.
}
\label{fig:performance}
\end{figure}

\paragraph{CPU Performance}
For the tests of the all-to-all interactions, we run \halmd{} (migration started at version 1.0.1) on a CPU using a set of 2,000 particles over 5,000 integration steps.
We employ a 6-core Intel Xeon W-2133 (Cascade Lake) workstation processor with 32\,GB of RAM running Debian 11 as operating system.
The LLVM-based Intel Compiler (icpx) version 2023.1.0 is used to compile both the sequential \Cpp{} version as well as the SYCL version of the code.
In both cases, we let the compiler optimize with \code{-O3} for the processor's microarchitecture (\code{-mtune}), enable vectorization and allow for fast math optimization, while checking that computational results were correct.
In CPU benchmarks, we use double floating-point precision.
Since the SYCL version allows for parallel execution, we scale the number of threads by setting the \code{DPCPP\_CPU\_NUM\_CUS} environment variable accordingly.

With a single thread, we observed a remarkable 5.1-fold speedup of the SYCL version over the original sequential code.
Further analysis using Intel's APS tool showed that the sequential version, although compiled with the according flags, could not be vectorized by the compiler and only 15.8\% of the DP FLOPs are performed using 128-bit AVX SIMD instructions.
On the other hand, for the SYCL version, written in a data-parallel way, the compiler vectorizes all of the relevant code for AVX-512, which is a plausible cause of the very different execution times obtained.

The speedups obtained for parallel execution on the CPU are displayed in Fig.~\ref{fig:performance}a using two different baselines:
the original sequential version and the single-thread SYCL version.
Using all six cores of the CPU, the SYCL code shows a 21.9-fold speedup compared to the original one.
We point out that this large value is a positive side effect of the migration efforts, which originally intended to port the code to different GPUs; a tuning of the original sequential code was out of this work's scope.
Figure~\ref{fig:performance}a shows also the parallel efficiency of the SYCL version, and
we conclude that it scales well and nearly linearly, delivering a parallel efficiency of 0.84 on average.

\paragraph{GPU Performance for All-to-All Interaction}
For the actual portability and performance assessment we employed three different GPUs from two vendors:
An Nvidia A40 RTX 48GB, an Nvidia A100 SXM4 80GB, and an AMD MI-210.
On the software side, Nvidia's CUDA SDK version 11.0.2 is employed for the native CUDA version on both Nvidia GPUs.
For compilation of the SYCL version, we use the open source Intel LLVM compiler 2022.12 with the Nvidia CUDA toolkit 11.7 as backend.
(The slight difference in CUDA toolkit versions are due to an incompatibility
of the CUB library versions shipped with \halmd{} and with newer CUDA SDKs.
On the other hand, CUDA SDK 11.0.2 is not supported as a backend for SYCL with this version of LLVM.)
For the Nvidia GPUs, the code was compiled for compute capability 8.0. 
When compiling for the AMD GPU, we used ROCm 5.0 targeting the \code{gfx90a} architecture;
the usage of OpenSYCL~\cite{DBLP:conf/iwocl/AlpaySWH22} (hipSYCL) was not considered within this work.
In the MD setup, the particle count remained at 2,000 whereas the number of integration steps was increased to 20,000.

As an outcome, we could execute the SYCL port of \halmd{} on all considered GPUs.
We also note that the SYCL code successfully runs on integrated (Gen9.5) and prototypical (ATS) Intel GPUs, but their performance is not comparable to the dedicated devices discussed here and is therefore not of interest for the present study.
The kernel execution times in Fig.~\ref{fig:performance}b show that the SYCL version is a about 1.3 times faster on the Nvidia devices than their CUDA counterpart.
In addition to that, the code runs also on AMD GPU devices, thus fulfilling the goal of portability.
Since no HIP version was available at the time of writing, a performance comparison with SYCL was not possible.

\paragraph{Truncated Force Performance Study}
Following to these initial and promising performance assessments, more realistic experiments employing truncated pair forces were conducted.
To this end, we restricted the study to execution on GPUs and increased the system size to 100,000 particles and the number of simulation time steps to 50,000.
We successfully executed the same code on all three considered GPUs and also successfully verified the computational results.
Thus, the goal of code portability was again reached.

However, contrary to the all-to-all interaction, the SYCL version of the truncated forces runs up to 1.8 times slower than the CUDA version on the A100 GPU as shown in Figure~\ref{fig:performance}c.
Similar observations were made on the A40 GPU.
A breakdown of the performance reveals that both the forces calculation and the update the of the Verlet neighbor lists, the SYCL versions (see Section~\ref{sec:halmd}) experience a slowdown of 1.8 and 1.5\texttimes, while having an approximate share of 60\% and 30\% on the runtime.
The share varies with the simulation temperature parameter (see Figure~\ref{fig:performance}c) as a higher temperature parameter causes more frequent neighbor list update due to increased particle movement.

Care has been taken that the launch configurations of the SYCL and CUDA kernels are identical, such that both versions can make use of the same number of GPU resources.
In addition, the compilation flags permit the usage of FMA operations.

For closer investigation of the kernels, the Nvidia \emph{ncu} profiler was used.
However the usually unnamed lambda kernel functions had to be converted to named lambdas because the filtering of the profiler could not properly cope with long type info names.
Nevertheless, the profiling process revealed major differences.
First, we note that for both kernels, the SYCL version does not use FMA operations as efficiently as the CUDA version.
We attribute this to the generated code as the manually migrated source code essentially performs the same operations.
In case for the force computation, the lower FMA usage contributes to a reduction of the compute throughput by 68\% compared to the CUDA version.
The neighbor list-update is similarly affected (64\% lower throughput).

The profiling also revealed that the SYCL version of the force calculation experiences warp execution stalls 1.8\texttimes{} more often than the CUDA version.
We attribute this to the lack of texture memory support in the SYCL version (see Section~\ref{sec:halmd:migration}).
Instead, the has to be fetched from the global memory resulting in additional load operations and no benefit of loading the packed data from texture memory can be made.
Further, the usage of global memory can reduce performance by a factor of 1.47 on an A100 GPU\footnote{\url{https://github.com/ricsonc/linear_vs_texture_memory_cuda}} compared to the usage of texture memory.
Overall, both the compiler effects and the employed memory types can be identified as reasons the reduced SYCL performance.

\section{Summary and Conclusion}
For the \halmd{}, we have successfully demonstrated that SYCL enables code portability across platforms.
Although tool support exists, different migration strategies must be considered to migrate complex codes to SYCL which may even include manual migration of code.
For the particular application, it was shown that such a migration may even result in better execution on a contemporary CPU with SIMD units.
More importantly, the successful migration also enables to execution of \halmd{} on different GPU platforms of different vendors.

With respect to performance, we note that this actually depends on the considered application kernel.
While we observed, that one can gain performance boost of up to 1.3\texttimes{} for straight-forward computations, we also experienced performance degradations for SYCL which we attributed to both memory type effects and quality of the generated code.
It will therefore be of strong interest to see whether the performance will improve with further adoption of SYCL within the community and increasing feature support by SYCL implementations.


\bibliography{references}

\begin{thebibliography}{AGG20}

\bibitem[Ab15]{Abraham:S2015}
Abraham, Mark~James et~al.: {GROMACS}: High performance molecular simulations
  through multi-level parallelism from laptops to supercomputers.
\newblock {SoftwareX}, 1-2:19--25, 2015.

\bibitem[AGG20]{Anderson:CMS2020}
Anderson, Joshua~A.; Glaser, Jens; Glotzer, Sharon~C.: {HOOMD}-blue: A {P}ython
  package for high-performance molecular dynamics and hard particle {M}onte
  {C}arlo simulations.
\newblock Comput. Mater. Sci., 173:109363, 2020.

\bibitem[Al22]{DBLP:conf/iwocl/AlpaySWH22}
Alpay, Aksel; Soproni, B{\'{a}}lint; W{\"{u}}nsche, Holger; Heuveline, Vincent:
  Exploring the possibility of a hipSYCL-based implementation of oneAPI.
\newblock In: IWOCL'22: Int. Workshop on OpenCL, Bristol, United Kingdom, May
  10 - 12, 2022.
\newblock {ACM}, pp. 10:1--10:12, 2022.

\bibitem[APL21]{Alekseenko_2021}
Alekseenko, Andrey; P{\'{a}}ll, Szil{\'{a}}rd; Lindahl, Erik: Experiences With
  Adding {SYCL} Support to {GROMACS}.
\newblock In: Int. Workshop on {OpenCL}.
\newblock {ACM}, April 2021.

\bibitem[AT17]{AllenTildesley:Simulation}
Allen, M.~P.; Tildesley, D.~J.: Computer simulation of liquids.
\newblock Oxford University Press, Oxford, 2 edition, 2017.

\bibitem[Ba17]{Bailey:SP2017}
Bailey, Nicholas et~al.: {RUMD}: A general purpose molecular dynamics package
  optimized to utilize {GPU} hardware down to a few thousand particles.
\newblock {SciPost} Phys., 3(6), 2017.

\bibitem[Br09]{Brooks:JCC2009}
Brooks, B.~R. et~al.: {CHARMM}: The Biomolecular Simulation Program.
\newblock J. Comp. Chem., 30:1545--1615, 2009.

\bibitem[CH09]{Colberg2009}
Colberg, Peter~H.; Höfling, Felix: Highly accelerated simulations of glassy
  dynamics using GPUs: caveats on limited floating-point precision.
\newblock Comput. Phys. Commun., 2009.

\bibitem[CH16]{Chaudhuri:2016}
Chaudhuri, Pinaki; Horbach, Jürgen: Structural inhomogeneities in glasses via
  cavitation.
\newblock Phys. Rev. B, 94:094203, 2016.

\bibitem[Co22]{DBLP:conf/iwbbio/CostanzoRSNP22}
Costanzo, Manuel et~al.: Migrating {CUDA} to oneAPI: {A} Smith-Waterman Case
  Study.
\newblock In (Rojas, Ignacio et~al., eds): Bioinf. Biomed. Eng. - 9th Int. Work
  Conf., {IWBBIO} 2022, Maspalomas, Gran Canaria, Spain, June 27-30, 2022,
  Proceedings, Part {II}.
\newblock volume 13347 of Lecture Notes Comput. Sci. Springer, pp. 103--116,
  2022.

\bibitem[CS20]{DBLP:conf/ipps/ChristgauS20}
Christgau, Steffen; Steinke, Thomas: Porting a Legacy {CUDA} Stencil Code to
  oneAPI.
\newblock In: 2020 {IEEE} Int. Parallel and Distributed Processing Symposium
  Workshops, {IPDPSW} 2020, New Orleans, LA, USA, May 18-22, 2020.
\newblock {IEEE}, pp. 359--367, 2020.

\bibitem[{cu}23]{cuda-wrapper}
{cuda\_wrapper}: , A lightweight {C++11} wrapper of the {CUDA} {API},
  2007--2023.
\newblock \url{https://github.com/halmd-org/cuda-wrapper}.

\bibitem[Ea17]{Eastman:PCB2017}
Eastman, Peter et~al.: {OpenMM} 7: Rapid development of high performance
  algorithms for molecular dynamics.
\newblock {PLOS} Comput. Biol., 13(7):e1005659, 2017.

\bibitem[Eb20]{EbrahimiViand:JCP2020}
Ebrahimi~Viand, Roya; Höf{}ling, Felix; Klein, Rupert; Delle~Site, Luigi:
  Communication: {T}heory and Simulation of Open Systems out of Equilibrium.
\newblock J. Chem. Phys., 153(10):101102, 2020.

\bibitem[FT22]{DBLP:conf/waccpd-ws/FortenberryT22}
Fortenberry, Anna; Tomov, Stanimire: Extending {MAGMA} Portability with OneAPI.
\newblock In: 9th Workshop on Accelerator Programming Using Directives,
  WACCPD@SC 2022, Dallas, TX, USA, November 13-18, 2022.
\newblock {IEEE}, pp. 22--31, 2022.

\bibitem[HD15]{Hoefling:2015}
H{\"o}f{}ling, Felix; Dietrich, Siegfried: Enhanced wavelength-dependent
  surface tension of liquid--vapour interfaces.
\newblock EPL (Europhys. Lett.), 109(4):46002, 2015.

\bibitem[H{\"o}16]{Hoeft:Thesis}
H{\"o}ft, Nicolas: Computer Simulations of Phase Behavior and Adsorption
  Kinetics in Metal-Organic Frameworks ({MOFs}).
\newblock PhD thesis, Univ. Düsseldorf, 2016.

\bibitem[H{\"o}23]{HALMD}
H{\"o}fling, F. et~al.: , {H}ighly {A}ccelerated {L}arge-scale {M}olecular
  {D}ynamics package, 2007--2023.
\newblock \url{https://halmd.org}.

\bibitem[Ne16]{osti_1332474}
Neely, J.~R.: DOE Centers of Excellence Performance Portability Meeting.
\newblock April 2016.

\bibitem[Ph20]{Phillips:JCP2020}
Phillips, James~C. et~al.: Scalable molecular dynamics on {CPU} and {GPU}
  architectures with {NAMD}.
\newblock J. Chem. Phys., 153(4):044130, 2020.

\bibitem[RDH16]{Roy:2016}
Roy, Sutapa; Dietrich, Siegfried; Höf{}ling, Felix: Structure and dynamics of
  binary liquid mixtures near their continuous demixing transitions.
\newblock J. Chem. Phys., 145(13):134505, 2016.

\bibitem[RHK22]{std:sycl:2020r5}
Rovatsou, Maria; Howes, Lee; Keryell, Ronan: , {SYCL 2020 Specification}, May
  2022.

\bibitem[RSF13]{Salomon-Ferrer:WCMS2013}
R.~Salomon-Ferrer, D.A.~Case, R.C.~Walker: An overview of the {A}mber
  biomolecular simulation package.
\newblock {WIRE}s Comput. Mol. Sci., 3:198--210, 2013.
\newblock \url{https://ambermd.org/GPUSupport.php}.

\bibitem[Sa23]{DBLP:conf/supercomputer/SakiotisAPRTZ23}
Sakiotis, Ioannis et~al.: Porting Numerical Integration Codes from {CUDA} to
  oneAPI: {A} Case Study.
\newblock In (Bhatele, Abhinav; Hammond, Jeff~R.; Baboulin, Marc; Kruse,
  Carola, eds): High Performance Computing - 38th Int. Conf., {ISC} High
  Performance 2023, Hamburg, Germany, May 21-25, 2023, Proceedings.
\newblock volume 13948 of Lecture Notes Comput. Sci. Springer, pp. 339--358,
  2023.

\bibitem[St20]{Straube:CP2020}
Straube, A.~V.; Kowalik, B.~G.; Netz, R.~R.; Höf{}ling, F.: Rapid onset of
  molecular friction in liquids bridging between the atomistic and hydrodynamic
  pictures.
\newblock Commun. Phys., 3:126, 2020.

\bibitem[Th22]{Thompson:CPC2022}
Thompson, Aidan~P. et~al.: {LAMMPS} - a flexible simulation tool for
  particle-based materials modeling at the atomic, meso, and continuum scales.
\newblock Comput. Phys. Commun., 271:108171, February 2022.

\bibitem[To23]{Tovey:JC2023}
Tovey, Samuel et~al.: {MDSuite}: comprehensive post-processing tool for
  particle simulations.
\newblock J. Cheminf., 15(1), 2023.

\bibitem[We19]{weik19a}
Weik, Florian et~al.: {ESPResSo} 4.0 -- An Extensible Software Package for
  Simulating Soft Matter Systems.
\newblock Eur. Phys. J. Special Topics, 227(14):1789--1816, 2019.

\end{thebibliography}

\end{document}